\documentclass[
reprint,
superscriptaddress,
amsmath,amssymb,
aps,
]{revtex4-1}

\usepackage{graphicx,epsf}
\usepackage{float}
\usepackage{xcolor}

\begin{document}

\title{Normal and inverted hysteresis in perovskite solar cells}

\author{George Alexandru Nemnes}
\email{nemnes@solid.fizica.unibuc.ro}
\affiliation{University of Bucharest, Faculty of Physics, Materials and Devices for Electronics and Optoelectronics Research Center, 077125 Magurele-Ilfov, Romania}
\affiliation{Horia Hulubei National Institute for Physics and Nuclear Engineering, 077126 Magurele-Ilfov, Romania}
\author{Cristina Besleaga}
\affiliation{National Institute of Materials Physics, Magurele 077125, Ilfov, Romania}
\author{Viorica Stancu}
\affiliation{National Institute of Materials Physics, Magurele 077125, Ilfov, Romania}
\author{Lucia Nicoleta Leonat}
\affiliation{National Institute of Materials Physics, Magurele 077125, Ilfov, Romania}
\author{Lucian Pintilie}
\affiliation{National Institute of Materials Physics, Magurele 077125, Ilfov, Romania}
\author{Kristinn Torfason}
\affiliation{School of Science and Engineering, Reykjavik University, Menntavegur 1, IS-101 Reykjavik, Iceland}
\author{Marjan Ilkov}
\affiliation{School of Science and Engineering, Reykjavik University, Menntavegur 1, IS-101 Reykjavik, Iceland}
\affiliation{Icelandic Heart Association, Holtasmari 1, IS-201 Kopavogur, Iceland}
\author{Andrei Manolescu}
\affiliation{School of Science and Engineering, Reykjavik University, Menntavegur 1, IS-101 Reykjavik, Iceland}
\author{Ioana Pintilie}
\email{ioana@infim.ro}
\affiliation{National Institute of Materials Physics, Magurele 077125, Ilfov, Romania}

\begin{abstract}
Hysteretic effects are investigated in perovskite solar cells in the
standard FTO/TiO$_2$/CH$_3$NH$_3$PbI$_{3-x}$Cl$_x$/spiro-OMeTAD/Au
configuration.  We report normal (NH) and inverted hysteresis (IH)
in the J-V characteristics occurring for the same device structure,
the behavior strictly depending on the pre-poling bias. NH typically
appears at pre-poling biases larger than the open circuit bias, while
pronounced IH occurs for negative bias pre-poling. The transition
from NH to IH is marked by a intermediate mixed hysteresis behavior
characterized by a crossing point in the J-V characteristics. The
measured J-V characteristics are explained quantitatively by the dynamic
electrical model (DEM). Furthermore, the influence of the bias scan rate
on the NH/IH hysteresis is discussed based on the time evolution of the
non-linear polarization.  Introducing a three step measurement protocol,
which includes stabilization, pre-poling and measurement, we put forward
the difficulties and possible solutions for a correct PCE evaluation.
\end{abstract}

\maketitle

\section{Introduction}

The dynamic hysteresis phenomena observed in the J-V characteristics
of perovskite solar cells (PSCs) fuels an ongoing debate about its
origin, accurate determination of the photoconversion efficiency
(PCE), and cell stability \cite{snaith,dualeh,elumalai,egger}. Several
mechanisms have been proposed to explain the hysteretic behavior:
ion migration \cite{tress15,bastiani,eames,zhao,chenbo}, charge
trapping and de-trapping \cite{shao,jixian}, ferroelectric
polarization \cite{wei,chen,frost1,frost2}, capacitive effects
\cite{perez,almora,sanchez}, charge accumulation at the interfaces
\cite{wubo,sepalage}, or unbalanced distribution of electrons and
holes \cite{bergmann}. It is already established that the hysteresis is
influenced by measurement settings such as the bias scan rate and range
\cite{tress15}, but also by the cell pre-conditioning by voltage poling
\cite{cao} and light soaking \cite{AENM:AENM201500279}.

In the typical hysteresis the reverse-scan current, i.e. measured
by reducing the voltage from open circuit bias ($V_{oc}$) to short
circuit, is larger than in the forward scan. During the reverse
scan the cell behaves like a capacitor, releasing charge in the
external circuit, and hence an excess of current is obtained, while during the
forward scan the opposite situation occurs.  We call this behavior
{\it normal hysteresis} (NH). Recently  an {\it inverted hysteresis}
(IH) was reported \cite{tress16} in mixed-cation mixed-halide PSCs,
and attributed to charge extraction barriers. It was further shown that
methylammonium-lead-iodide (MAPbI$_3$) PSCs with TiO$_2$ layer covered
by a thin Al$_2$O$_3$ insulating shell also exhibits IH. In other studies
the IH was explained in terms of ionic accumulation \cite{jacobs},
and observed in aged samples with Mo/Ag counter electrodes \cite{besleaga}.

In this paper we report for the first time the presence of both
NH and IH in the same sample, strictly depending on the applied
pre-poling bias ($V_{pol}$): NH is obtained for $V_{pol}> V_{oc}$ and
IH for $V_{pol}<0$. For $0<V_{pol}<V_{oc}$ the hysteresis loop has a
crossing point where the forward and reverse characteristics meet. The
experimental data are well reproduced by the dynamic electrical model 
which is summarized below.  
Then, we indicate a three-step J-V measurement protocol which is
essential for a meaningful evaluation of the hysteretic effects. We
discuss the pre-poling bias conditions for the realization of NH, IH
and intermediate mixed hysteresis.  In connection with the measurement
protocol, by controlling the switching between NH and IH, we provide
further insight into PSC operation but also a systematic assessment of the
dynamic hysteresis, which is crucial for a proper evaluation of the PCE.



\begin{figure}
\includegraphics[width=8.cm]{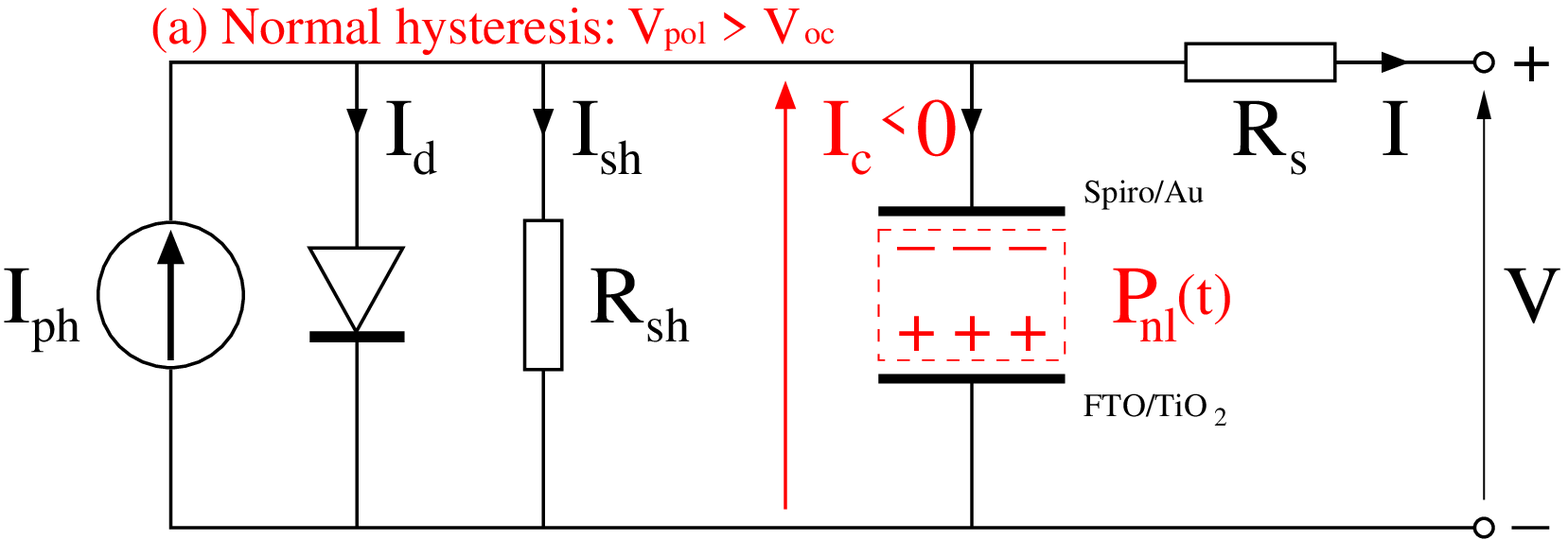}
\includegraphics[width=8.cm]{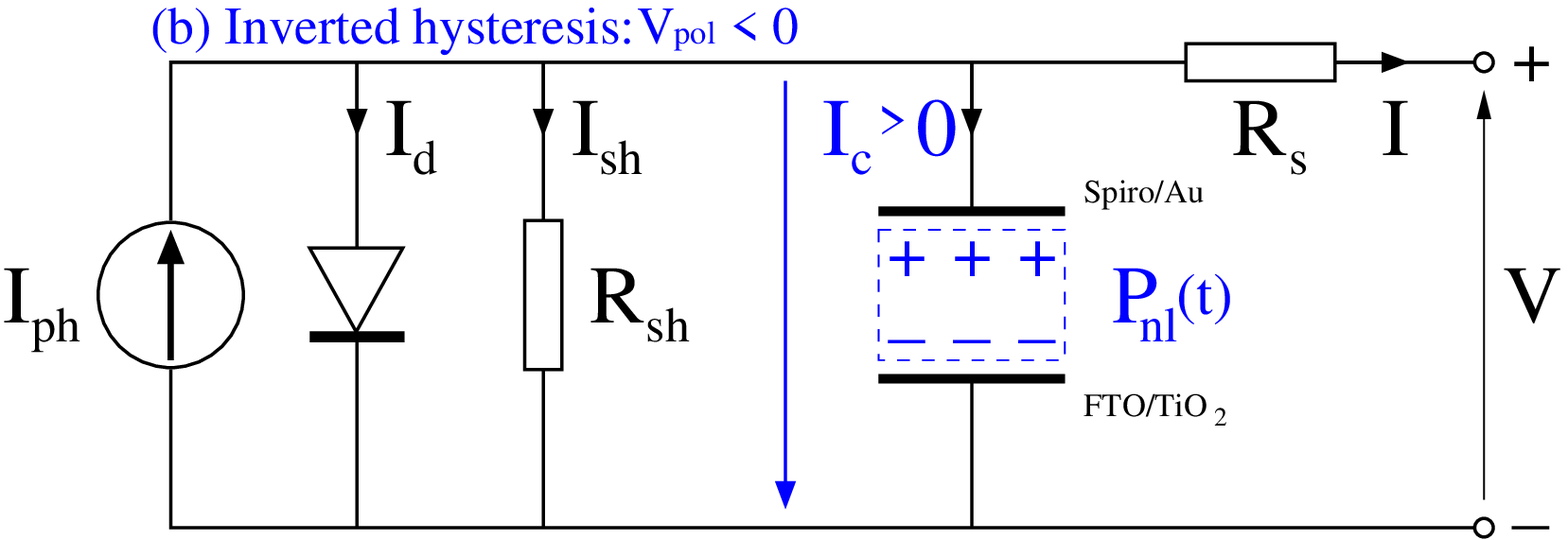}
\caption{Schematics of the dynamic electrical model for two different poling conditions: (a) $V_{pol}>V_{oc}$ (NH) and (b) $V_{pol}<0$ (IH) induce opposite polarization orientations. The sign of $I_c\sim\partial P_{nl}/\partial t$ is indicated in both cases for a reverse scan starting from $V_{oc}$, after the poling step.   
}
\label{DEM}
\end{figure}

\section{The dynamic electrical model (DEM)}

Several models have been considered for the dynamic hysteresis:
Initially a drift-diffusion approach has been formulated, accounting
for ionic migration, electronic charge traps, and recombination centers
\cite{reenen}, or coupled charge carriers and defect mediated ion
motion including bias scan rate and pre-poling \cite{richardson}. Then,
an equivalent circuit model with constant capacitance was proposed
\cite{seki}. The DEM, recently formulated by our group
\cite{nemnes1}, is also based on an equivalent circuit, but includes a 
capacitor accounting for non-linear polarization effects 
as depicted in Fig.\ \ref{DEM}.  The key assumption
is that the slow process governing the time evolution of the non-linear
polarization $P_{nl}$ can be described by a specific relaxation time $\tau$.  
Since $\tau$ is typically of the order of seconds DEM is compatible
with ion migration and accumulation at the interfaces resulting in a
surface polarization of Maxwell-Wagner-Sillars type.
The time dependent J-V characteristics is determined by the coupled 
differential equations:
\begin{eqnarray}
\label{Idiffeq2}
-R_s C_0 \frac{\partial I}{\partial t} &=&
      I_s \left( e^{\frac{q(V + I R_s)}{n k_B T}}-1\right) \nonumber \\
&+& \left(\frac{R_s}{R_{sh}} + 1 \right) I \nonumber \\
&+& \frac{V}{R_{sh}} + C_0 \frac{\partial V}{\partial t}
    + {\mathcal A}\frac{\partial P_{nl}}{\partial t} - I_{ph}\ , 
\end{eqnarray}
\begin{equation}
\frac{\partial P_{nl}}{\partial t} = \frac{P_{nl,\infty}(U_c(t)) - P_{nl}(t)}{\tau}\ ,
\label{dPdt}
\end{equation}
with initial conditions $I(t=0)=I_0$ and $P_{nl}(t=0)=P_0$, which
determines the NH or IH behavior.  The current density $J$ is defined as the 
ratio between the current intensity $I$ and the device active area ${\mathcal A}$. 
The elements of the circuit are: the
series resistance $R_s$, the shunt resistance $R_{sh}$, the diode ideality
factor $n$, the diode saturation current $I_s$, the photogenerated current
$I_{ph}$ and the geometric capacitance $C_0$. In contrast to the standard
model with a constant capacitance, the time dependence of the non-linear
polarization $P_{nl}$ obeys Eq.\ (\ref{dPdt}), with a constant relaxation
time $\tau$, whereas the equilibrium polarization $P_{nl,\infty}=(U_c/V_{oc})
P_\infty$ depends on the voltage on the non-linear capacitor, $U_c=V + I R_s$, 
where $V$ is the external applied bias. The constant polarization
$P_\infty$ corresponds to $U_c=V_{oc}$ in the steady
state. The relaxation time $\tau$ accounts for the decay of the initial
polarization $P_0$ and also for the evolution of the polarization
when the applied bias is changed. 
In fact, as it will be shown in the following, the current component 
$I_c^{(nl)}= {\mathcal A} \times \partial P_{nl}/ \partial t$ has a crucial role in producing 
the hysteretic effects, representing a significant part of the measured current I:
\begin{equation}
I = I_{ph} - I_d - I_{sh} - I_c,
\label{I}
\end{equation}
where $I_d$ is the diode current, $I_{sh}$ accounts for recombinations and 
$I_c = I_c^{(l)} + I_c^{(nl)}$, with typically negligible linear component $I_c^{(l)}$
due to the small geometrical capacitance $C_0$. 
However, depending on the pre-poling conditions $I_c^{(nl)}$ can reach magnitudes similar 
to the measured current \cite{nemnes1}.

Another version of DEM has been subsequently proposed by Ravishankar
et. al.  \cite{bisquert} in terms of a relaxation kinetic constant
determined by ion displacement, by reformulating Eq. (\ref{dPdt}) for
the surface polarization voltage [Eq. (6) therein].  Both versions of
DEM consistently reproduced experimental NH, including the dependence
on the bias scan rate and the current overshoot in the reverse scan
observed in several experimental studies \cite{tress15,meloni,sanchez},
reproduced only in part by drift-diffusion based models \cite{richardson}.
The overshoot is consistent with a relaxation time of the polarization 
or charge accumulated at interfaces \cite{nemnes1,bisquert}
at the beginning of the reverse scan measurement, indicating a 
relation between the pre-poling bias and the initial polarization.
Furthermore, the dependence of the hysteresis on the bias scan rate is
consistently described by the DEM: the hysteresis is diminished at very
low and very high bias scan rates, being significant at intermediate ones;
importantly, the short circuit current is enhanced with the increase of
the bias scan rate.

It is worth noting that a standard {\it constant} capacitance model
cannot explain the current overshoot, nor can account for the diminished
hysteresis at very high scan rates, unless a vanishing dielectric constant
for rapidly varying fields is assumed.  Moreover, it predicts only the
normal hysteresis behavior.

\section{Measurement protocol}

As already mentioned in other papers\cite{christians,zimmermann}, in order to 
obtain reproducible measurements, it is important to adopt a well defined 
measurement protocol.
We consider three steps: 1 - the {\it stabilization} of the open circuit bias $V_{oc}$, 
2 - the solar cell {\it pre-conditioning} phase by pre-poling under 1 sun illumination, 
and 3 - the actual J-V {\it measurement} consisting of a single reverse-forward bias scan starting from the open circuit bias to short-circuit and back. Before each pre-conditioning phase, the solar cell is kept in open circuit under illumination until the steady state is achieved, corresponding to stabilized $V_{oc}$ and zero current. In this way we reinitialize as far as possible the current state of the solar cell, which may be affected in successive measurements. We choose the open circuit condition for the stabilization phase, instead of the short-circuit, to avoid as much as possible the solar cell deterioration. 
Pre-poling with illumination is achieved by applying a constant bias $V_{pol}$ 
for a time $t_{pol}$. For $V_{pol}>V_{oc}$ we call the sample over-polarized, 
for $0<V_{pol}<V_{oc}$ under-polarized, and for $V_{pol}<0$ reversely polarized. 
Regarding the actual measurement, instead of performing individual forward and reverse bias scans, the single step reverse-forward bias scan ensures a better control over the sample pre-poling, eliminating potential pre-poling differences in the two separate scans.

\section{Results and discussion}


\begin{figure}[t]
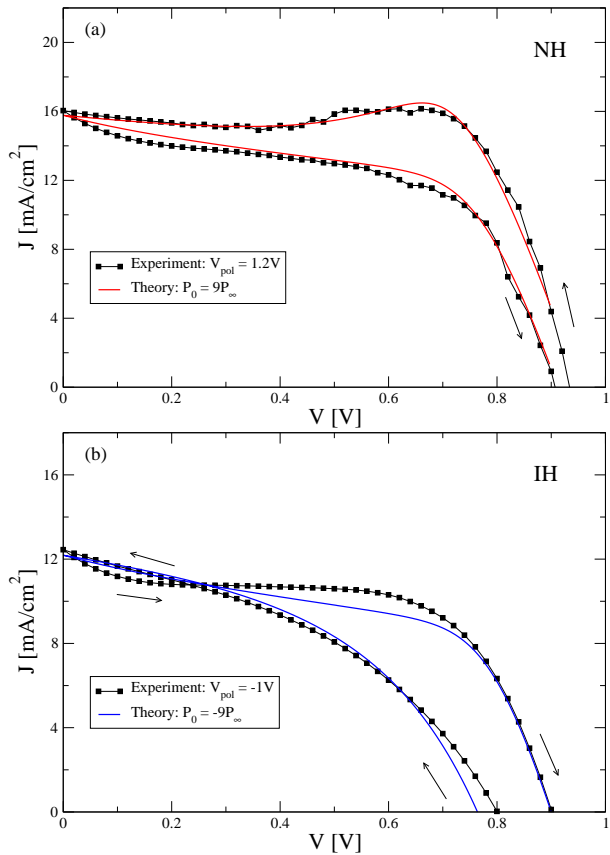

\centering
\includegraphics[width=8.cm]{figure2a}
\includegraphics[width=8.cm]{figure2b}
\caption{
(a) Calibration of the simulated J-V characteristics using the measured data of a pre-poled sample, at $V_{pol}=1.2$V for $t_{pol}=$30 s, exhibiting NH, at a bias scan rate of 20 mV/s. The corresponding initial polarization is $P_0=9P_\infty$. (b) J-V characteristics of the same sample pre-poled at $V_{pol}=-1V$ and the resulting simulated J-V characteristics, obtained with the same parameters, while changing $P_0=-9P_\infty$ and a smaller $I_{ph}=1.11$ mA. The IH is present for $0.23V<V<V_{oc}$, while NH is obtained for $V<0.23$V.
}
\label{NH_IH}
\end{figure}

{\it Results and discussion.}
In the following we discuss the NH and IH induced
by different pre-poling regimes, the measurements being consistently
reproduced by the DEM calculations. The samples are prepared in the standard FTO/TiO$_2$/CH$_3$NH$_3$PbI$_{3-x}$Cl$_x$/spiro-OMeTAD/Au configuration, with $x=0.4$, \cite{besleaga} as detailed in the Supporting Information (SI).
More details, including the physical properties and time evolution of similar samples, are described in Ref.\ \cite{besleaga}.
First, we calibrate the equivalent circuit elements with a measured
J-V characteristics showing NH, performed at a bias
scan rate of 20 mV/s, with positive pre-poling bias $V_{pol}=1.2V$
for $t_{pol}=$30 s. With this choice, the current overshoot is present
in the reverse characteristics and this marked feature enables a more
accurate extraction of the model parameters \cite{nemnes1,bisquert}.
The simulated J-V characteristics of the over-polarized sample is depicted
in Fig.\ \ref{NH_IH}(a) matching closely the detailed features of the
experimental data.  The obtained DEM parameters are: the photogenerated
current $I_{ph}=1.42$ mA, the series resistance $R_{s}=95\Omega$, the
shunt resistance $R_{sh}=3 k\Omega$, the diode ideality factor $n=1.53$
and saturation current $I_s=0.1$ pA. The relative dielectric permittivity
$\epsilon_r=100$ yields a quite small geometrical capacitance $C_0$
and its influence over the J-V characteristics is negligible in
comparison with the non-linear polarization component. The steady state
polarization at the open circuit is $P_\infty=18$ mC/cm$^2$, corresponding
to a stabilized $V_{oc}=0.9$ V. The relaxation time $\tau=9$ s fits well with the
preliminary indication about the magnitude of the characteristic
times involved.  The assumed initial polarization $P_0=9P_\infty$
corresponds to the pre-poling stage performed under illumination at
1.2 V for 30 s. The device active area is ${\mathcal A}=0.09$ cm$^2$,
which is also the illuminated area.

By switching the initial polarization $P_0$ to negative values, J-V characteristics with mixed or completely inverted hysteresis are observed. It is quite remarkable that changing only one parameter, $P_0=-9P_\infty$, the shape of the inverted hysteresis resembles a measured characteristics with $V_{pol}=-1$V for 30 s at the same bias scan rate, apart from the diminished short circuit current obtained in the measurement. Taking into account $I_{ph}=1.11$ mA the simulated J-V characteristics overlaps quite well with the experimental data, as indicated in Fig.\ \ref{NH_IH}(b). Our test case shows a mostly inverted hysteresis, with a crossing point in the forward and reverse characteristics at $\sim$0.23 V. 
The decrease of $I_{sc}$ at negative pre-poling was systematically observed in repeated measurements, starting from lower $V_{pol}$ values to higher ones and in reverse order, which will be discussed later.
Apart from this poling effect, a temporary aging of the cell is observed after repeated bias cycling, manifested by an overall decrease of the current in both forward and reverse characteristics. Typically, the samples recover in one day in dark and inert atmosphere conditions.

\begin{figure}[t]
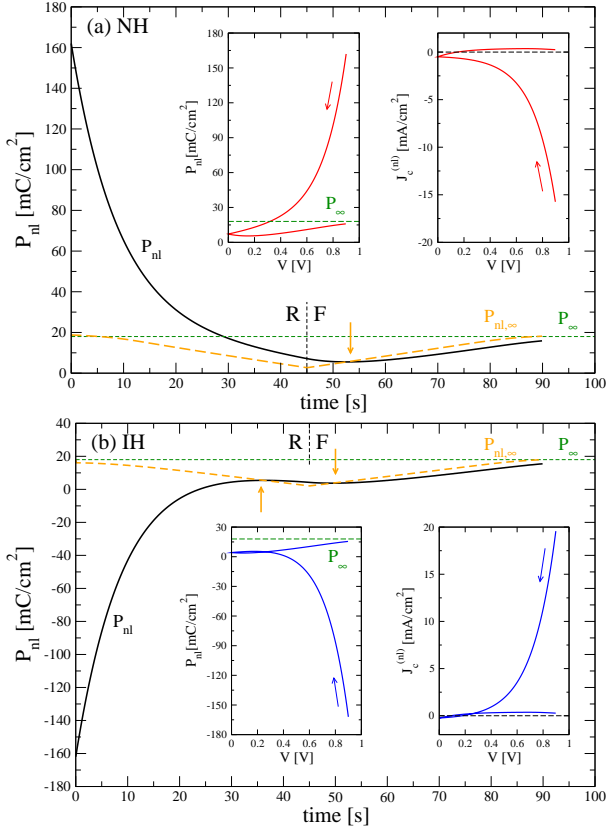

\centering
\includegraphics[width=8.cm]{figure3a}
\includegraphics[width=8.cm]{figure3b}
\caption{
Time evolution of the non-linear polarization $P_{nl}$ for NH (a) and IH (b), corresponding to J-V characteristics shown in Fig.\ \ref{NH_IH}. The vertical arrows mark the crossing points between the bias dependent steady state polarization $P_{nl,\infty}$ and the current polarization $P_{nl}$. The insets show the bias dependence of $P_{nl}$ and $J_c^{(nl)}=\partial P_{nl}/\partial t$. The vertical dashed lines mark the separation between the reverse and forward regimes. 
}
\label{Pnl_t}
\end{figure}

Both hysteretic effects can be explained by analyzing the time evolution of the non-linear polarization $P_{nl}(t)$, depicted in Fig.\ \ref{Pnl_t}. In the case of NH, $P_{nl}(t=0)=P_0\gg P_\infty$ is positive and decays rather fast in a time interval of the order of $\tau$ in the beginning of the reverse scan. Subsequently, $P_{nl}$ follows the bias dependent steady state polarization $P_{nl,\infty}(U_c(t))$ [Eq.\ (\ref{dPdt})] as the bias is further decreased in reverse and then increased in the forward regime. On the other hand, IH is produced by negative poling, yielding a negative initial polarization, $P_0<0$. The measurement, however, starts from $V_{oc}$ and $P_{nl}$ is driven into positive region on the reverse scan, reaching a maximum, and then decreasing towards the zero-bias state. As $\tau$ is here smaller than the total time of the reverse scan, the behavior of $P_{nl}$ in the forward scan is similar for both IH and NH. The condition $P_{nl,\infty}=P_{nl}$ is correlated either with a minimum or a maximum for $P_{nl}(t)$. As already indicated, the $J_c^{(nl)}$ current component is quantitatively significant and plays a determinant role in the features on the J-V characteristics. Depending on its sign the value of the measured current can be enhanced or diminished according to Eq.\ (\ref{I}), influencing the determined PCE value.

\begin{figure}[t]
\centering
\includegraphics[width=9.cm]{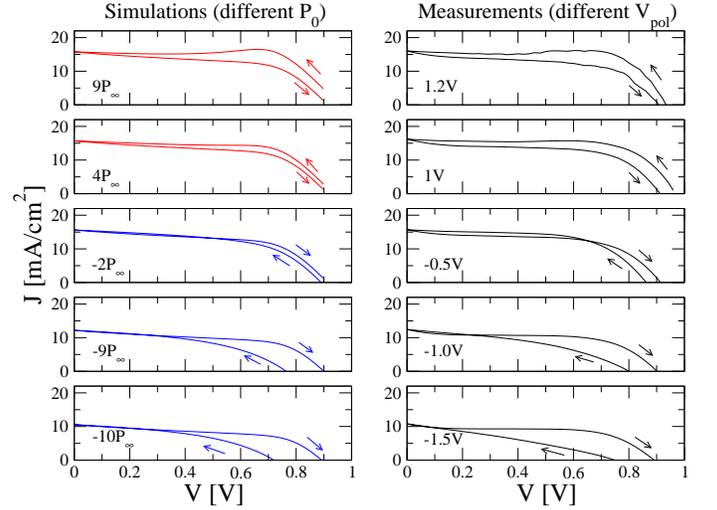}
\caption{Normal and inverted hysteresis, obtained by changing the pre-poling bias $V_{pol}$ (measurements, right panel) and, correspondingly, the initial polarization $P_0$ (simulations, left panel), for a bias scan rate of 20 mV/s.  
}
\label{nhih_prepol}
\end{figure}

To further investigate the transition between NH and IH, we measured
J-V characteristics with several pre-poling voltages, between -1.5V to
1.2V, as shown in Fig.\ \ref{nhih_prepol} together with the calculated
J-V characteristics with suitably chosen initial polarizations $P_0$.
A crossing point between the reverse and forward characteristics appears
and it moves systematically towards lower biases as $P_0$ (or $V_{pol}$)
decreases.  This behavior was theoretically predicted before, in Ref.\
\cite{nemnes1}, as an under-polarization regime ($0<P_0<P_\infty$). On
the other hand, the crossing was observed experimentally \cite{besleaga}
on aged samples with Ag counter electrodes, using a forward-reverse scan
protocol starting from -0.1 V, which may be the effect of a small negative
bias pre-poling, although not intentionally performed. The crossing
of separately measured forward and reverse characteristics was also
observed \cite{bisquert}.  As already indicated, a particular feature is
the decrease of the $I_{sc}$ found in the measured J-V characteristics, as
negative poling increases in absolute value. Based on this experimental
observation and the fact that $R_{sh}$ is relatively large, which
implies $I_{sc}\approx I_{ph}$, and assuming a relaxation time independent 
on the poling bias, we adjust the $I_{ph}$
current to 1.11 mA ($P_0=-9P_\infty$) and 0.96 mA ($P_0=-10P_\infty$)
for $V_{pol}$ = -1 V and -1.5 V, respectively.
This may be explained by iodine migration during the poling time,
as negative iodine ions tend to accumulate at the TiO$_2$ mesoporous layer
and the induced electric field reduces electron collection during the subsequent 
measurement \cite{jacobs}. By contrast, when positive poling is applied the electric field
induced by ions is favorable to electron extraction and $I_{sc}$ does not
change significantly.

\begin{figure}[t]
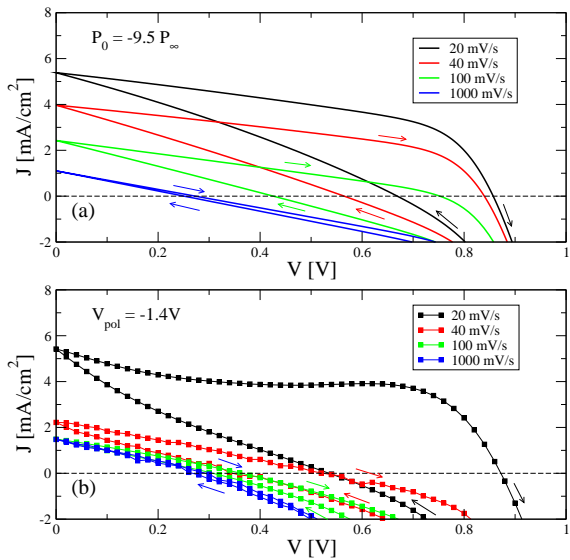

\centering
\includegraphics[width=7.5cm]{figure5a}\\
\includegraphics[width=7.5cm]{figure5b}
\caption{Inverted hysteresis under varying the bias scan rate: (a) simulated J-V characteristics for bias scan rates in the range 20 - 1000 mV/s, with $P_0=-9.5P_\infty$; (b) measured J-V characteristics, showing the decreasing of $I_{sc}$ with increasing the bias scan rate.  
}
\label{ih_scan_rate}
\end{figure}

We focus next on the bias scan rate dependence of the IH. For the NH 
we obtained a consistent
correspondence between the measurements and DEM \cite{nemnes1}:
the NH is widest at intermediate bias scan rates and $I_{sc}$
increases at higher rates. This behavior was explained
in terms of time evolution of non-linear polarization $P_{nl}$ and the
current component $J_c^{(nl)}=\partial P_{nl} /\partial t$ introduced in
Eq.\ (\ref{dPdt}), as also shown (for comparison to IH) in Fig.\
\ref{sNHIHsim} of SI.  The simulated and measured J-V characteristics showing IH are
presented in Fig.\ \ref{ih_scan_rate}.  In the experiment we considered
$V_{pol}=-1.4$ V, which corresponds in simulations to $P_0=-9.5P_\infty$. 
 At this stage the sample
is slightly aged, as one may notice from the 20 mV/s scan, having a smaller
$I_{sc}$ compared to bias pre-poling measurements. We therefore consider
$I_{ph}=0.6$ mA and an increased $\tau=30$ s and reproduce the sequence of
experimental J-V characteristics with different bias scan rates.  As in
the case of NH at very fast bias scans, the hysteresis amplitude is
diminished. However, the behavior of $I_{sc}$ is opposite: it becomes
smaller as the scan rate increases.  At very fast bias scan rates
there is little variation in $P_{nl}$, as it remains close to $P_0$,
and $\partial P_{nl}/\partial V$ becomes small in both forward and
reverse scans, reducing thus the hysteresis of both NH and IH.  
In contrast, at short-circuit, the calculated current $J_c^{(nl)}=
\partial P_{nl} /\partial t$ has a larger magnitude for NH at
higher bias scan rates, and the opposite occurs for IH (Fig. \ref{sNHIHsim}). This accounts
for both measured and calculated decrease in $I_{sc}$ in IH conditions 
(Figs. \ref{sNHIHpoling} and \ref{sNHIHscanrate}).  
We also note that $\partial P_{nl} /\partial t < 0$ for NH and $\partial P_{nl}
/\partial t > 0$ for IH, which is the reason for the crossing of the
forward and reverse characteristics.  We therefore conclude that the
pre-poling bias is a determinant switching factor from NH to IH, which
become two forms of hysteresis previously considered independent.

\begin{figure}[t]
\centering
\includegraphics[width=9.cm]{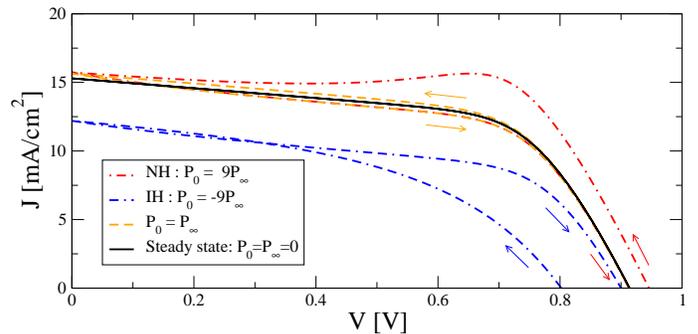}
\caption{Analyzing the PCEs under different poling conditions in comparison with the reference value  obtained from the static J-V characteristics: NH:10.75\%, IH:4.47\%, $P_\infty$:8.54\% in reverse bias scan and
NH:8.24\%, IH:6.12\% $P_\infty$:8.23\% in forward scan; the steady state yields the PCE value of 8.44\%. 
}
\label{PCE}
\end{figure}

The dynamic hysteresis creates difficulties for an accurate
determination of the solar cell PCE. 
Figure\ \ref{PCE} shows NH ($P_0=9P_\infty$), IH ($P_0=-9P_\infty$) cases 
and a minimal polarization case ($P_0=P_\infty$), for a 20 mV/s bias scan rate,
and the steady state solution ($P_\infty=0$), in the bias range $[0,V_{oc}]$. 
The PCE varies drastically
in forward and reverse scans of both NH and IH, with a large deviation in
the reverse scans compared to the steady state PCE: 27\% overestimation
for NH and 48\% underestimation for IH. 
In the forward scan of the NH we find a deviation of only 2.3\%, while 
for IH we obtain a 28\% smaller PCE.
However in the minimal polarization
case, where the reverse scan starts at $V_{oc}$ determined before
the actual measurement, only a small hysteresis is found,  
as $P_\infty\neq0$. This case provides the most accurate description of the steady
state, with a deviation of less than 3\% for both forward and reverse scans. 
Decreasing the bias scan rate would reduce the hysteresis loop
even further. However, in practical measurements, for a reasonable
measurement time, it is very important to determine firstly $V_{oc}$, in
order to avoid over-polarization. Secondly, another recommendation within
the proposed measurement protocol is taking a continuous reverse-forward
cycle, which is essential for removing an unintended initial polarization,
e.g. by applying a reverse scan ending into negative voltage followed by
a separate forward scan after an arbitrary time interval. A waiting time
between reverse and forward bias scans influence significantly the second
(forward) scan, as it is further detailed in Fig.\ \ref{tw} (SI), for an IH test case. 
Therefore we
emphasize the importance of a careful consideration of the cell dynamic
polarization. Our three-step protocol allows a systematic evaluation of
such hysteretic phenomena, but also provides a realistic determination 
of the PCE as the initial polarization is minimized.

\section{Conclusions}

To conclude, we investigated the dynamic hysteresis of standard perovskite
solar cells with controlled pre-conditioning. By varying the
pre-poling bias, both normal and inverted hysteresis are found for the
same sample and the measured J-V characteristics are closely reproduced.  
The main findings of the present study are:

\begin{enumerate}

\item The NH and IH can be obtained by pre-poling the solar cell with bias
$V_{pol}>V_{oc}$ and $V_{pol}<0$, respectively, and a mixed hysteresis is typically
observed for $0<V_{pol}<V_{oc}$.

\item The DEM reproduces the two hysteretic behaviors by changing only
the initial polarization parameter, which is related to experimental
quantities $V_{pol}$ and $t_{pol}$.

\item At higher bias scan rates the short circuit current $I_{sc}$ is
enhanced for NH and lowered for IH, consistently explained by DEM.

\item The observed NH/IH behavior reveals once again the importance of
controlling the solar cell pre-conditioning, particularly the bias
pre-poling; the state induced by a previous measurement may inadvertently
affect the next one; the designed three-step protocol, including {\it
stabilization}, {\it pre-poling} and {\it measurement} is afforded for
the reliability and reproducibility of the experimental data, being
especially important for a correct PCE determination.

\item The presence of both NH and IH unifies apparently contradictory results
reported in recent papers. Our results indicate the presence of both
normal and the inverted hysteresis in the same device if the
required pre-poling conditions can be achieved.\\

\end{enumerate}

\begin{acknowledgments}
The research leading to these results has received funding from EEA Financial Mechanism 2009-2014 under the project contract no 8SEE/30.06.2014. 
\end{acknowledgments}

\bibliography{manuscript}


\onecolumngrid

\appendix* 
\section{Supporting Information}

\renewcommand{\thefigure}{S\arabic{figure}}
\setcounter{figure}{0}

{\bf 1. Device fabrication and characterization.}\\


\hspace*{0.cm}{\bf a) Device fabrication}\\
\hspace*{1.2cm}We fabricated perovskite solar cells by successive spin-coating deposition of TiO2 thin and meso-porous layers, CH$_3$NH$_3$PbI$_{3-x}$Cl$_x$ mixed halide perovskite and spiro-OMeTAD on commercial glass/FTO substrate (Solaronix TCO22-7). The compact TiO$_2$ blocking layer of 150 nm thickness is obtained from bis(acetylacetonate) solution (Aldrich) and annealing at 450 $^\circ$C for 30 min. The mesoporous TiO$_2$ film (mp-TiO$_2$), composed of 20 nm size particles is deposited using a solution of TiO$_2$ commercial paste (Solaronix Ti-Nanoxide N/SP) diluted in ethanol (1:3, weight ratio). The mp-TiO$_2$ structures were first annealed at 150 $^\circ$C for 5 min and then crystallized at 500 $^\circ$C for 1 h. The CH$_3$NH$_3$PbI$_{3-x}$Cl$_x$ mixt halide perovskite is fabricated using the principle of one-step method \cite{Ann2015}: a precursor solution containing 369 mg lead iodide, 56 mg lead chloride, 78 mg methyl sulfoxide, 159 mg methyl ammonium iodide (Dysol) and 600 mg dimethylformamide, homogenized for one hour, was spin coated with 2000 rpm/25 s. After 9 s from the start of the spin cycle 150 $\mu$l of diethyl ether was dripped on top of the layer, enabling the solvent extraction process to take place. The final perovskite film is obtained after annealing at 65 $^\circ$C for 1 min and 100 $^\circ$C for 2 min. All the above described processes were performed in normal laboratory conditions, at 24 $^\circ$C and humidity between 30\% and 40\%. The spiro-OMeTAD was spin-coated at 1500 rpm for 30 s, in N$_2$ enriched atmosphere, at 24 $^\circ$C and humidity less than 10\%. The solution used for this deposition was obtained by mixing 80 mg spiro-OMeTAD (Borun Chemical), 28 $\mu$l 4-tert-butylpyridine and 18 $\mu$l of bis(trifluoromethane)sulfonimide lithium salt in acetonitrile solution (520 mg ml$^{-1}$). Gold counter electrodes of 0.09 cm$^2$ area have been deposited by magnetron RF sputtering.\\ 


\hspace*{0.cm}{\bf b) Electrical characterization}\\
\hspace*{1.2cm}The Photocurrent--Voltage (J--V) characteristics were measured using an Oriel VeraSol-2 Class AAA LED Solar Simulator having AM 1.5 filters and a Keithley 2400 Source Meter. The irradiation intensity was 100 mW/cm$^2$, calibrated by a Newport standard silicon solar cell 91150. The 1 Sun illumination has been performed through a rectangular aperture of 3$\times$3 mm$^2$ size, of a geometry identical with that of Au counter electrodes.\\ 

\vspace*{0.5cm}



\newpage

{\bf 2. Simulations and measurements revealing normal hysteresis (NH) or inverted hysteresis (IH), for different bias scan rates and prepoling biases.}\\

\vspace*{-0.0cm}
{\bf a) Simulated NH and IH for different bias scan rates}\\

\hspace*{4.3cm}{\bf NH} \hspace*{5.8cm}{\bf IH}\\ \vspace*{-0.5cm}
\begin{figure}[h]
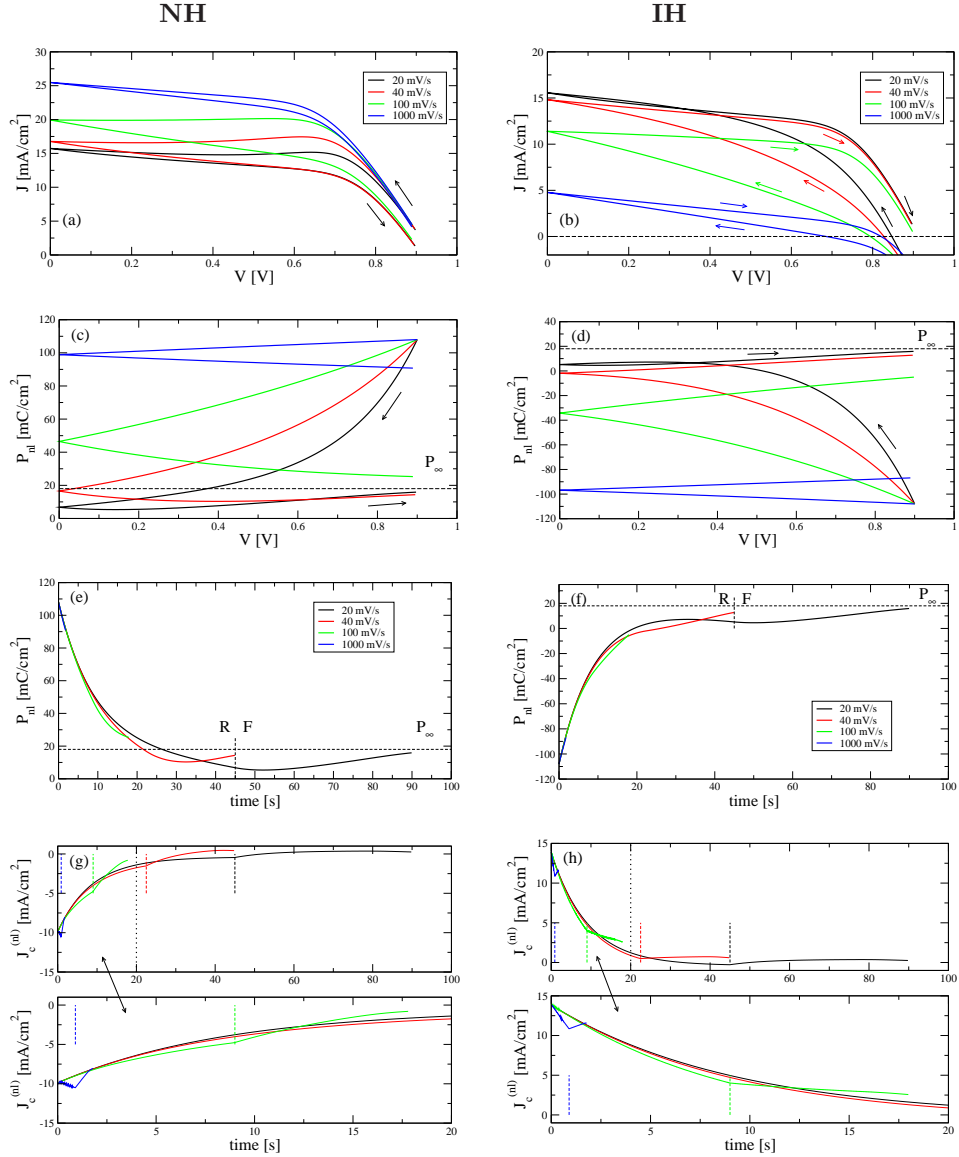

\centering
\includegraphics[width=6.cm]{JV_NH_scan_rate_simulated_2}\hspace*{0.5cm}
\includegraphics[width=6.cm]{JV_IH_scan_rate_simulated_2}\\ \vspace*{0.35cm}
\includegraphics[width=6.cm]{Pnl_V_NH_scan_rate_2}\hspace*{0.5cm}
\includegraphics[width=6.cm]{Pnl_V_IH_scan_rate_2}\\ \vspace*{0.35cm}
\includegraphics[width=6.cm]{Pnl_t_NH_scan_rate_2}\hspace*{0.5cm}
\includegraphics[width=6.cm]{Pnl_t_IH_scan_rate_2}\\ \vspace*{0.35cm}
\includegraphics[width=6.cm]{Jcnl_NH_2}\hspace*{0.5cm}
\includegraphics[width=6.cm]{Jcnl_IH_2}\\
\caption{Simulated reverse-forward NH and IH behavior for different bias scan rates, with pre-poling $P_0 = 6P_\infty$ and $P_0 = -6P_\infty$, while all other parameters are the same as the ones used in Fig. 2(a). (a,b) J-V characteristics.
(c,d) Non-linear polarization $P_{nl}$ as a function of applied bias. 
Time evolution of $P_{nl}$ (e,f) and $J_c^{(nl)}$ (g,h). 
At short-circuit (marked by dashed lines) $J_c^{(nl)}$ is increasing in absolute value with the scan rate.}
\label{sNHIHsim}
\end{figure}

\newpage

\hspace*{1.5cm}{\bf b) Experimental NH and IH induced by bias pre-poling} 

\begin{figure}[h]
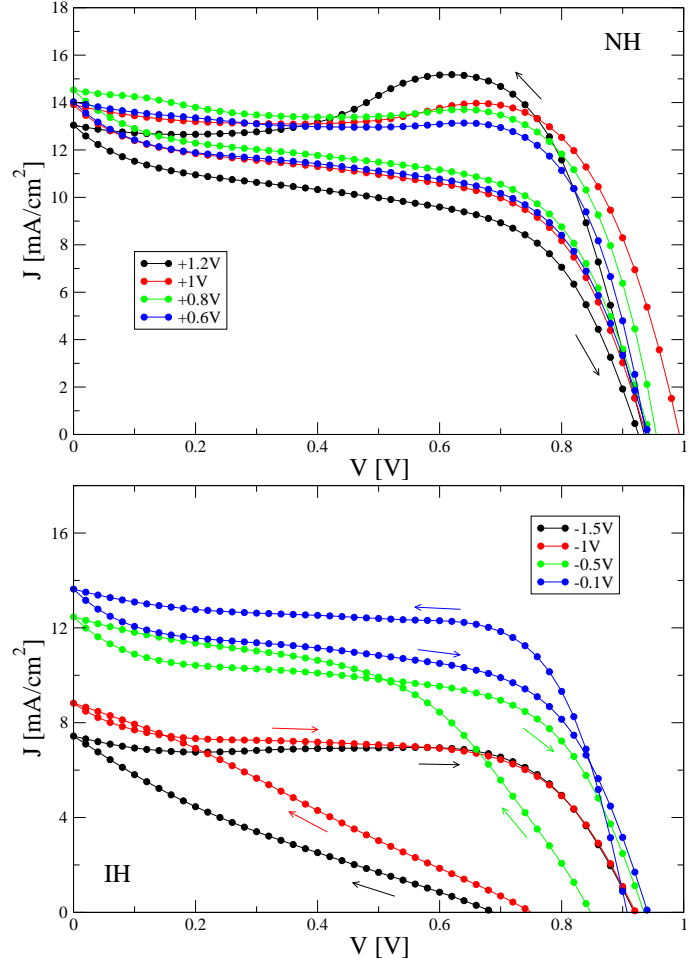

\centering
\includegraphics[width=9.cm]{experimental_JV_NH_prepoling}\\
\includegraphics[width=9.cm]{experimental_JV_IH_prepoling}
\caption{Experimental J-V characteristics showing NH and IH under pre-poling conditions: NH is obtained for positive pre-poling biases $V_{pol}=$ 0.6V, 0.8V, 1.0V, 1.2V, IH is obtained for $V_{pol}=$ -1.5V, while the mixed hysteresis characterized by a crossing point is found for $V_{pol}=$ -0.1V, -0.5V, -1V. The poling time is in each case $t_{pol}=30$ s.}
\label{sNHIHpoling}
\end{figure}

\newpage

\hspace*{0cm}{\bf c) Experimental NH and IH - influence of the bias scan rate} 

\begin{figure}[h]
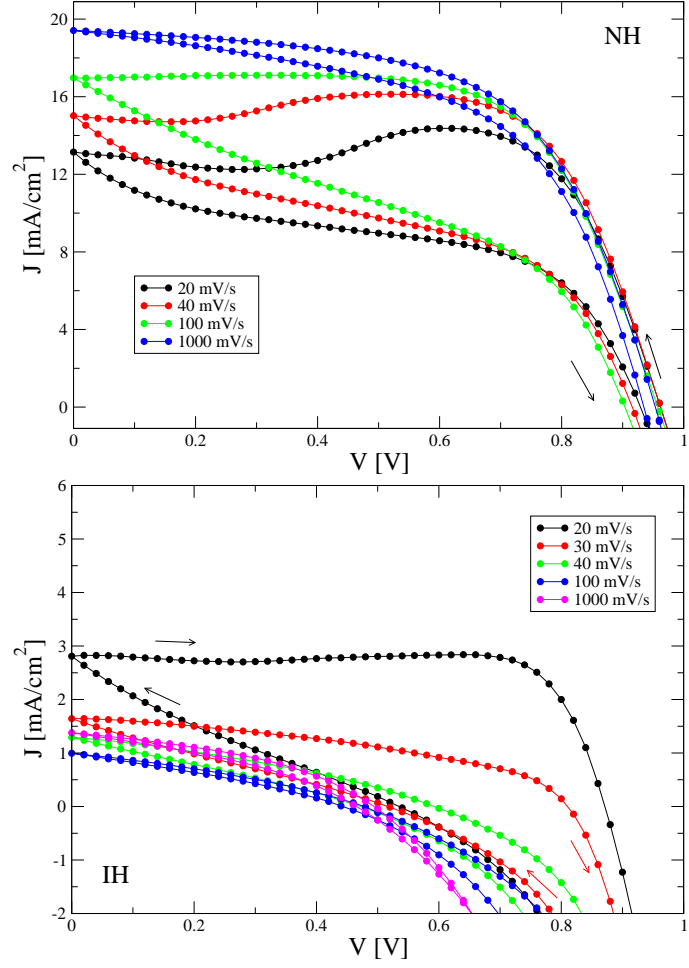

\centering
\includegraphics[width=9.cm]{experimental_JV_NH_scanrate}\\
\includegraphics[width=9.cm]{experimental_JV_IH_scanrate}
\caption{Experimental J-V characteristics showing NH and IH for different scan rates, in the range 20 - 1000 mV/s. The samples exhibiting NH are pre-poled at $V_{pol}=$ 1.2V, while IH characteristics are obtained for $V_{pol}=$ -1.5V, with $t_{pol}=30$ s.}
\label{sNHIHscanrate}
\end{figure}

\newpage

{\bf 3. Influence of the waiting time at short-circuit, between reverse and forward\\ \hspace*{0.5cm} scan, for the case of inverted hysteresis.}

\begin{figure}[h]
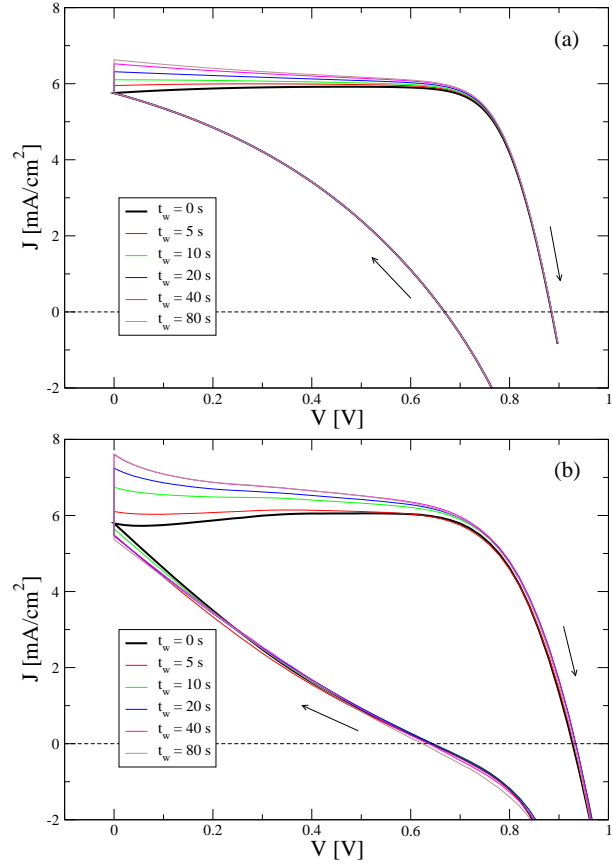

\centering
\includegraphics[width=8.cm]{JV_tw_simulation_2}\\
\includegraphics[width=8.cm]{JV_tw_experiment}
\caption{The influence of the waiting time $t_w$ spent at open circuit, between the reverse and forward scans (20 mV/s): simulation (a) and experiment (b). In the simulation, the relaxation time considered is $\tau = 20$ s, while the experimental pre-poling conditions are $V_{pol}=-1.5$V and $t_{pol}=30$ s. Other parameters which are modified for this particular sample with respect to calibration set are: $R_{sh}=30$ k$\Omega$ and $I_{ph}=0.6$ mA, while the initial polarization is $P_0=-12P_\infty$. Beyond $t_w=40$ s, the J-V characteristics tend to overlap.}
\label{tw}
\end{figure}

\end{document}